\documentclass[twocolumn,showpacs,aps,prl,groupedaddress]{revtex4}
\usepackage{epsfig,amssymb,amsmath}

\def\labell#1{\label{#1}}
\def\section#1{{\em #1:--- }}

\def\togli#1{}

\usepackage{color}

\begin{document}
\title{ Sequential projective measurements for channel decoding}

\author{Seth Lloyd$^1$, Vittorio Giovannetti$^2$, Lorenzo Maccone$^3$}
\affiliation{$^1$Dept.~of Mechanical Engineering, Massachusetts
  Institute of
  Technology, Cambridge, MA 02139, USA\\
  $^2$ NEST, Scuola Normale Superiore and Istituto Nanoscienze-CNR, 
piazza dei Cavalieri 7, I-56126 Pisa, Italy 
  \\
$^3$Dip.~Fisica ``A.~Volta'',  INFN Sez.~Pavia, Universit\`a di Pavia, via Bassi 6,
I-27100 Pavia, Italy}

\begin{abstract}
  We study the transmission of classical information in quantum
  channels. We present a decoding procedure that is very simple but
  still achieves the channel capacity. It is used to give an
  alternative straightforward proof that the classical capacity is
  given by the regularized Holevo bound.  This procedure uses only
  projective measurements and is based on successive ``yes''/``no''
  tests only.
\end{abstract}
\maketitle 

According to quantum information theory, to transfer classical signals
we must encode them into the states of quantum information carriers,
transmit these through the (possibly noisy) communication channel, and
then decode the information at the channel output \cite{BENSHOR}.
Frequently, even if no entanglement between successive information
carriers is employed in the encoding or is generated by the channel, a
joint measurement procedure is necessary (e.g.~see \cite{fuchs}) to
achieve the capacity of the communication line, i.e.  the maximum
transmission rate per channel use \cite{BENSHOR}. This is clear from
the original proofs \cite{HOL,SCHU} that the classical channel
capacity is provided by the regularization of the Holevo bound
\cite{BOUND}: these proofs employ a decoding procedure based on
detection schemes (the `Pretty-Good-Measurement' or its
variants~\cite{TYSON,MOCHO,BEL,BAN,UTS,ELD,HAUS,BARN,MONT,FIU,HAYDEN,KHOLEVO}).
Alternative decoding schemes were also derived in \cite{WINTER} (with
a combinatorial approach) and in \cite{OGAWA,HAYA,HAYA1} (with an 
application of quantum hypothesis testing, which was introduced in
this context in \cite{VERDU}).  Here we present a simple decoding
procedure which uses only dichotomic projective measurements, but
which is nonetheless able to achieve the channel capacity. 

The main idea is that even if the possible alphabet states (i.e.~the
states of a single information carrier) are not orthogonal at the
output of the channel, the codewords composed of a long sequence of
alphabet states approach orthogonality asymptotically, as the number
of letters in each codeword goes to infinity. Thus, one can
sequentially test whether each codeword is at the output of the
channel. When one gets the answer ``yes'', the probability of error is
small (as the other codewords have little overlap with the tested
one). When one gets the answer ``no'', the state has been ruined very
little and can be still employed to further test for the other
codewords. To reduce the accumulation of errors during a long sequence
of tests that yield ``no'' answers, every time a ``no'' is obtained,
we have to project the state back to the space that contains the
typical output of the channel.  Summarizing, the procedure is: 1.~test
whether the channel output is the first codeword; 2.~if ``yes'', we
are done, if ``no'', then project the system into the typical subspace
and abort with an error if the projection fails; 3.~Repeat the above
procedure for all the other codewords until we get a ``yes'' (or abort
with an error if we test all of them without getting ``yes'').  4.~In
the end, we identified the codeword that was sent or we had to abort.

We start reviewing some basic notions on typicality. Then, we prove
that the above procedure achieves the classical capacity of the
channel. An alternative (more formal) proof that refers to this same
method is presented in \cite{noteslong} by using a decoding strategy
in which the ``yes/no'' measurements discriminates only among the
typical subspaces of the codewords.  An application of our scheme to
communication over Gaussian bosonic channels~\cite{WERNER} will be
presented in Ref.~\cite{sihui}.

\section{Definitions and review}
For notational simplicity we will consider codewords composed of
unentangled states. For general channels, entangled codewords must be
used to achieve capacity~\cite{HASTINGS}, but the extension of our
theory to this case is straightforward (replacing the Holevo bound
with its regularized version).

Consider a quantum channel that is fed with a letter $j$ from a
classical alphabet with probability $p_j$. The letter $j$ is encoded
into a state of the information carriers which is evolved by the
channel into an output $\rho_j = \sum_k p_{k|j} |k\rangle_j\langle
k|$, where $_j\langle k'|k\rangle_j = \delta_{k'k}$. Hence, the
average output is
 \begin{eqnarray}
  \rho = \sum_{j}  p_j \rho_j = \sum_{j,k} p_j p_{k|j} |k\rangle_j\langle k|
  = \sum_k p_k |k\rangle\langle k|\;,
\end{eqnarray} 
where $|k\rangle_j$ and $|k\rangle$ are the eigenvectors of the $j$th
output-alphabet density matrix and of the average output respectively. The
subtleties of quantum channel decoding arise because the $\rho_j$
typically commute neither with each other nor with $\rho$.  The
Holevo-Shumacher-Westmoreland (HSW) theorem \cite{HOL,SCHU} implies
that we can send classical information reliably down the channel at a
rate (bits per channel use) given by the Holevo quantity~\cite{BOUND} 
 \begin{eqnarray}\label{HOLEchi}
\chi \equiv 
S(\rho) - \sum_{j}  p_j S(\rho_j) \;, \label{chi}
\end{eqnarray}
where $S(\cdot)\equiv-{\rm Tr}[(\cdot) \log_2(\cdot) ]$ is the von
Neumann entropy.  This rate can be asymptotically attained in the
multi-channel uses scenario as $\lim_{n \rightarrow \infty} (\log_2
N_n)/n$, where a set ${\cal C}_n$ of ${N}_n$ codewords $\vec j =
(j_1,\cdots, j_n)$ formed by long sequences of the letters $j$ are
used to reliably transfer ${N}_n$ distinct classical messages.
Similarly to the Shannon random-coding theory \cite{COVER}, the
codewords $\vec j\in {\cal C}_n$ can be chosen at random among the
{\em typical sequences} generated by the probability $p_j$, in which
each letter $j$ of the alphabet occurs approximately $p_j n $ times.
As mentioned in the introduction, the HSW theorem uses the
`Pretty-Good-Measurement' procedure to decode the codewords of ${\cal
  C}_n$ at the output of the channel. We will now show that a sequence
of binary projective measurements suffices~\cite{BINARY}.

\section{Sequential measurements for channel decoding}The channel
output state $\rho_{\vec j} \equiv \rho_{j_1} \otimes \cdots \otimes
\rho_{j_n}$ associated to a generic typical sequence $\vec{j}=( j_1,
\cdots, j_n)$ possesses a {\em typical subspace} ${\cal H}_{\vec j}$
spanned by the vectors $|k_1\rangle_{j_1} \cdots |k_n\rangle_{j_n}
\equiv |\vec k\rangle_{\vec j}$, where $|k\rangle_j$ occurs
approximately $p_j p_{k|j} n = p_{jk} n$ times, e.g.~see
Ref.~\cite{HOL}.  The subspace ${\cal H}_{\vec j}$ has dimensions
$\sim 2^{n\sum_j p_j S(\rho_j)}$ independent of the input $\vec j\in
{\cal C}_n$.
%
Moreover, a typical output subspace ${\cal H}$ and a projector $P$
onto it exist such that, for any $\epsilon>0$ and sufficiently large
$n$
\begin{eqnarray} {\rm Tr} \; \bar{\rho}
 >
  1-\epsilon
\labell{ea}\;,
\end{eqnarray}
where $\bar\rho \equiv P \rho\otimes \cdots \otimes \rho P$ is the
projection of the $n$-output average density matrix onto ${\cal H}$.
Notice that ${\cal H}$ and the ${\cal H}_{\vec{j}}$'s in general
differ. Typicality for ${\cal H}$ implies that, for $\delta>0$ and
sufficiently large $n$, the eigenvalues $\lambda_i$ of $\bar \rho$ and
the dimension of ${\cal H}$ are bounded as \cite{SCHU,HOL}
\begin{eqnarray}
&&\lambda_i\leqslant 2^{-n(S(\rho) -
    \delta)}
\labell{eigv}\;,
\\
&&
\mbox{\# nonzero eigenvalues}=
{\rm dim}({\cal H})\leqslant
2^{n(S(\rho)+\delta)}
\labell{typc}\;.
\end{eqnarray}
Define then the operator
\begin{eqnarray}
  \tilde\rho  =P \big( \; \sum_{\vec j,\vec k\in typ}p_{\vec
    j}p_{\vec k|\vec j}\; |\vec k\rangle_{\vec j}\langle\vec k|\; \big) P\leqslant\bar\rho
\labell{tilde}\;,
\end{eqnarray}
where the inequality follows because the summation is only restricted
to the $\vec{j}$'s that are typical sequences of the classical source,
and to the states $|\vec{k}\rangle_{\vec{j}}$ which span the typical
subspace of the $\vec{j}$-th output. [Without these limitations, the
inequality would be replaced by an equality.] Consequently the maximum
eigenvalue of $\tilde\rho$ is no greater than that of $\bar\rho$ while
the number of nonzero eigenvalues of $\tilde\rho$ cannot be greater
than those of $\bar\rho$, i.e.~Eqs.~\eqref{ea}--\eqref{typc} 
apply also to $\tilde\rho$.

Now we come to our main result. To distinguish between the $N_n$ distinct codewords of ${\cal C}_n$, we
perform sequential von Neumann
measurements corresponding to projections onto the possible outputs
$|\vec k\rangle_{\vec j}$ to find the channel input (as shown in
\cite{noteslong} these can also be replaced by joint projectors on the
spaces ${\cal H}_{\vec{j}}$).  In between these measurements, we
perform von Neumann measurements that project onto the typical output
subspace ${\cal H}$.

We will show that as long as the rate at which we send information
down the channel is bounded above by the Holevo quantity (\ref{chi}),
these measurements identify the proper input to the channel with
probability one in the limit that the number of uses of the channel
goes to infinity.  That is, we send information down the channel at a
rate $R$ smaller than $\chi$, so that there are  $N_n\simeq 2^{nR}$ possible
randomly selected codewords $\vec j$ that could be sent down over $n$
uses.  Each codeword gives rise to~$\sim 2^{n\sum_j p_j S(\rho_j)}$
possible typical outputs $|\vec k\rangle_{\vec j}$. As always with
{Shannon}-like random coding arguments \cite{COVER}, our set of
possible outputs only occupy a fraction $2^{-n (\chi- R)}$ of the full
output space.  This sparseness of the actual outputs in the full space
is the key to obtaining asymptotic zero error probability: all our
error probabilities will scale as $2^{-n (\chi -R)}$.

The codeword sent down the channel is some typical sequence $\vec j$,
which yields some typical output $|\vec k\rangle_{\vec j}$ with
probability $p_{\vec k| \vec j}$.  We begin with a von Neumann
measurement corresponding to projectors $P, \openone-P$ to check whether the
output lies in the typical subspace ${\cal H}$.  From Eq.~(\ref{ea})
we can conclude that for any $\epsilon > 0$, for sufficiently large
$n$, this measurement yields the result ``yes'' with probability
larger than $1-\epsilon$.  We follow this with a binary projective
measurement with projectors
\begin{eqnarray}
 P_{\vec k_1| \vec j_1} \equiv  
|\vec k_1\rangle_{\vec j_1} \langle \vec k_1|  , \quad
\openone -  P_{\vec k_1| \vec j_1}, 
\labell{3}\;
\end{eqnarray}
to check whether the input was $\vec j_1$ and the output was $\vec
k_1$.  If this measurement yields the result ``yes'', we conclude that
the input was indeed $\vec j_1$.  Usually, however, this measurement
yields the result ``no''.  In this case, we perform another
measurement to check for typicality, and move on to a second trial
output state, e.g., $|\vec k_2\rangle_{\vec j_1}$.  If this
measurement yields the result ``yes'', we conclude that the input was
$\vec j_1$.  Usually, of course, the measurement yields the result
``no'', and so we project again and move on to a third trial output
state, $|\vec k_3\rangle_{\vec j_1}$ etc.  Having exhausted the
$O(2^{n\sum_k p_k S(\rho_k)})$ typical output states from the codeword
${\vec j_1}$, we turn to the typical output states from the input
${\vec j_2}$, then ${\vec j_3}$, and so on, moving through the $N_n\simeq 2^{nR}$
 codewords until we eventually find a match.  The
maximum number of measurements that must be performed is hence 
\begin{eqnarray} \label{max}
M\simeq  2^{nR} \; 
2^{n\sum_k p_k S(\rho_k)}\;.
\end{eqnarray}
 The probability amplitude
that after $m$ trials without finding the correct state, we find it at
the $m+1$'th trial can then be expressed as
\begin{eqnarray} {\cal A}_m({yes}) = {}_{\vec j}\langle {\vec
    k}| P ( \openone - P_{\ell_{m}} ) P \cdots P (\openone - P_{\ell_1}) P |\vec
  k\rangle_{\vec j}\;,
\labell{4a}\;
\end{eqnarray}
where for $q=1,\cdots, m$
 the operators $P_{\ell_q}$ represent the first  $m$ elements 
 $P_{\vec k_r| \vec j_s}$ that compose the decoding sequence of projectors.
 The error
probability $P_{err}(\vec{j}, \vec{k})$ of mistaking the vector
$|\vec{k}\rangle_{\vec{j}}$ can then be bounded by considering the
worst case scenario in which the codeword sent is the last one tested
in the sequence. Since this is the worst that can happen, $|{\cal
  A}_M(yes)|$ with  $m=M$, is the smallest possible, so that $P_{err}
(\vec{j}, \vec{k}) \leqslant 1 - {|{\cal A}_M(yes)|^2}$.  Recall that
the input codewords ${\vec j}$ are randomly selected from the set of
typical input sequences, and $\vec k$'s are typical output sequences.
Then, the average error probability for a randomly selected set of
input codewords can bounded as $\langle {P}_{err} \rangle\leqslant 1 -
\langle\left|{{\cal A}_M(yes)}\right|^2\rangle\leqslant 1 -
\left|\langle{{\cal A}_M(yes)}\rangle\right|^2$. Here $\langle
\cdots\rangle$ represents the average over all possible codewords of a
given selected codebook  ${\cal C}_n$  {\em and} the averaging over all possible
codebooks of codewords. The Cauchy-Swarz inequality $\langle{|{\cal
    A}_M(yes)|^2}\rangle \geqslant \left|\langle{{\cal
      A}_M(yes)}\rangle\right|^2$ was employed.  The last term can be
evaluated as
\begin{eqnarray}
     && \langle{{\cal A}_m(yes)}\rangle = \nonumber 
    \mbox{Tr}\Big[
    P\Big(\openone-\sum_{\ell_{m}} \pi_{\ell_m}
       P_{\ell_{m}}\Big)P\cdots \\
      \nonumber &&   \quad    P\Big(\openone-\sum_{\ell_1}
      \pi_{\ell_1} 
      P_{\ell_1}\Big)P\tilde\rho\Big]=
\mbox{Tr}\Big[(P-\tilde\rho)^m\;\tilde\rho\Big] 
\\ && \quad \qquad =\sum_{k=0}^m\left(\begin{matrix}m\cr
    k\end{matrix}\right)(-1)^k\; \mbox{Tr}\Big[\tilde\rho^{k+1}\Big]\labell{bbl} 
\;,
\end{eqnarray}
 where $\pi_\ell$ stands for the probability $p_{\vec j}p_{\vec k|\vec j}$, and where 
we used \eqref{tilde} and \eqref{3} to write
$\tilde\rho=\sum_\ell \pi_\ell PP_\ell P$.
To prove the optimality of our decoding, it is hence  sufficient to show that
$ \langle{{\cal A}_m(yes)}\rangle\sim 1$ even when the number $m$ of
measurements is equal to its maximum possible value $M$ of
Eq.~(\ref{max}).  Consider then Eqs.~\eqref{eigv} and \eqref{typc}
which imply the inequalities
\begin{eqnarray}
\mbox{Tr}\tilde\rho^j\leqslant\sum{}_{i=0}^{{\rm dim}({\cal
      H})}\lambda_i^j
\leqslant2^{n[S(1-j)+\delta(1+j)]}
\labell{bounds}\;.
\end{eqnarray}
Use this and Eq.~(\ref{ea}) to rewrite Eq.~(\ref{bbl}) as
\begin{eqnarray} && \langle{{\cal A}_m(yes)}\rangle
  \geqslant\mbox{Tr}\tilde\rho+ \sum_{k=1}^m\left(\begin{matrix}m\cr
      k\end{matrix}\right)(-1)^k\mbox{Tr}\Big[\tilde\rho^{k+1}\Big]
\labell{eq15}
\\\nonumber
&&\geqslant 1-\epsilon-\sum_{k=1}^m\left(\begin{matrix}m\cr
    k\end{matrix}\right)2^{n[-kS(\rho)+\delta(k+2)]}
=1-\epsilon-\gamma,
\end{eqnarray}
where $\gamma\equiv 2^{2n\delta}[(1+\zeta_n)^m-1]$, with
$\zeta_n=2^{n[-S(\rho)+\delta]}$. If $S(\rho)>\delta$, for
large $n$ we can write
\begin{eqnarray}
  (1+\zeta_n)^m-1\simeq 
   e^{m\zeta_n}-1\simeq m\zeta_n
\labell{lll}\;.
\end{eqnarray}
Hence, $\gamma$ is asymptotically negligible as long as
$2^{2n\delta}\; m\; \zeta_n$ is vanishing for $n\to\infty$. This yields the
constraint
\begin{eqnarray}
m \leqslant 2 ^{ n (S(\rho) - \delta)}\quad \mbox{ for all }m\;.
\end{eqnarray} 
In particular, it must hold for $M$, the largest value of $m$ given in
(\ref{max}). Imposing this, the decoding procedure yields a vanishing
error probability if the rate $R$ satisfies
\begin{eqnarray}
R < \chi -\delta\;,  \label{ff}
\end{eqnarray}
as required by the Holevo bound~\cite{BOUND}.

Summarizing, we have shown that under the condition~(\ref{ff}) the
average amplitude $ \langle{{\cal A}_m(yes)}\rangle$ of identifying
the correct codeword is asymptotically close to $1$ even in the worst
case in which we had to check over {\em all} the other codewords
$m=M$. This implies that the average probability of error in
identifying the codeword asymptotically vanishes. In other words, the
procedure works even when the measurements are chosen so that the
codeword sent is the last one tested in the sequence of tests. Note
that the same results presented here can be obtained also starting
from the direct calculation of the error probability \cite{noteslong}
(instead of using the probability amplitude).

We conclude by noting that from Eq.~\eqref{4a} one immediately sees
that the POVM $\{E_\ell\}$ relative to the global decoding procedure is
\begin{eqnarray}
  &&E_1=PP_1P;\ E_2=P(\openone-P_1)PP_2P(\openone-P_1)P;\nonumber
\\&&\nonumber
  E_\ell=P(\openone-P_1)P(\openone-P_2)P\cdots
  P(\openone-P_{\ell-1})P\\&&
  \times P_\ell
  \cdots(\openone-P_1)P;\ E_{0}=\openone-
  \sum{}_{\ell=1}^{M} E_\ell
\labell{mpovm}\;,
\end{eqnarray}
where $P_\ell$ is defined as in \eqref{3} and $E_0$ is the ``abort''
result.  We gave a simple realization of this POVM using sequential
``yes/no'' projections, but different realizations may be possible. It
is an alternative to the conventional Pretty-Good-Mea\-su\-re\-ment.
Note also that, with the exception of $E_{0}$, all the operators in
this POVM are simply projections onto pure states or on their
orthogonal complement.  Such sequence of projective measurements is
`asymptotically unentangling' in the sense that the output state
departs at most infinitesimally from its original separable form
throughout the entire decoding procedure. This clarifies that the role
of entanglement in the decoding is analogous to \cite{qnwe}: namely,
increasing the distinguishability of a multi-partite set of states
that are not orthogonal when considered by separate parties that do
not employ entanglement.

\section{Conclusions} 
Using projective measurements in a sequential fashion, we gave a new
proof that it is possible to attain the Holevo capacity when a noisy
quantum channel is used to transmit classical information.  Such
measurements provide an alternative to the usual
Pretty-Good-Measurements for channel decoding, and can be used in many
of the same situations.  In particular, an analogous procedure can be
used to decode channels that transmit quantum information, to approach
the coherent information limit~\cite{sl,sh,DEV}.  This follows simply
from the observation~\cite{DEV} that the transfer of quantum messages
over the channel can be formally treated as a transfer of classical
messages imposing an extra constraint of privacy in the signaling.

VG acknowledges P.~Hayden, A.S.~Holevo, K.~Matsumoto, J.~Tyson, M. M.
Wilde, and A.~Winter for comments and discussions. VG was supported
from the FIRB-IDEAS project, RBID08B3FM, and from Institut
Mittag-Leffler.  SL was supported by the WM Keck Foundation, DARPA,
NSF, and NEC.


\begin{references}
\bibitem{BINARY}Binary projective measurements have Kraus
operators $\Pi ,\openone-\Pi $ ($\Pi $ being a projector): the outputs
``yes'' and ``no'' correspond to $\Pi $ and $\openone-\Pi$
respectively.  The probability of each outcome is $p={\rm Tr}[\rho M]$
with $M=\Pi $ or $M=\openone-\Pi $, and the post-measurement state is
$M\rho M/p$.
\end{references}
\end{document}